\documentstyle[graphicx,pra,manuscript,aps]{revtex}
\tightenlines

\begin{document}

\title{Atomic Solitons in Optical Lattices}
\author{S.~P\"otting$^{1,2}$, P.~Meystre$^1$ and E.~M.~Wright$^1$ \\
        $^1$Optical Sciences Center, University of Arizona\\
        Tucson, AZ 85721, USA\\
        $^2$Max-Planck Institut f{\"u}r Quantenoptik\\
        D-85748 Garching, Germany}

\maketitle

\section{Introduction}
\label{introduction}

The experimental demonstration of Bose--Einstein condensation in
atomic vapors \cite{Ande95,Davi95,Brad95} has rapidly lead to
spectacular new advances in atom optics. In particular, it has
enabled its extension from the linear to the nonlinear regime,
very much like the laser lead to the development of nonlinear
optics in the 1960s. It is now well established that two-body
collisions play for matter waves a role analogous to that of a
Kerr nonlinear crystal in optics. In particular, it is known that
the nonlinear Schr\"odinger equation which describes the
condensate in the Hartree approximation supports soliton
solutions. For the case of repulsive interactions normally
encountered in BEC experiments, the simplest solutions are dark
solitons, that is, `dips' in the density profile of the
condensate. These dark solitons have been recently demonstrated in
two experiments \cite{Dens00,Burg99} which appear to be in good
agreement with the predictions of the Gross--Pitaevskii equation.

While very interesting from a fundamental physics point-of-view,
dark solitons would appear to be of limited potential for
applications such as atom interferometry, where it is desirable to
achieve the dispersionless transport of a spatially localized
ensemble of atoms, rather than a `hole'. In that case, bright
solitons are much more interesting. However, the problem is that
large condensates are necessarily associated with repulsive
interactions, for which bright solitons might seem impossible
since the nonlinearity cannot compensate for the kinetic energy
part (diffraction) in the atomic dynamics. While this is true for
atoms in free space, this is however not the case for atoms in
suitable potentials, eg. optical lattices. This is because in that
case, it is possible to tailor the dispersion relation of the
atoms in such a way that their effective mass becomes negative.
For such negative masses, a repulsive interaction is precisely
what is required to achieve soliton solutions. This result is
known from nonlinear optics, where such soliton solutions, called
gap solitons, have been predicted and demonstrated 
\cite{Chen87,Chri89,Ster94,Eggl96}.

The article is organized as follows: In Section \ref{NLAO} we
briefly review the basic formalism for nonlinear atom optics with
a focus on Bose condensed systems. This leads to the
Gross--Pitaevskii equation, the nonlinear mean-field equation of
motion for the atomic condensate. In Section \ref{variety} we
introduce the bright and dark soliton solutions of these equations
and motivate the concept of gap solitons. In Section
\ref{lattices} we then describe a specific model for gap solitons
in a spinor Bose--Einstein condensate. We demonstrate how to launch
and control these solitons in Section \ref{control}. Finally,
Section \ref{summary} is a summary and conclusion.

\section{Nonlinear atom optics}
\label{NLAO}
%
Early experiments and theories concerning atom optics considered
the low-density regime where atom-atom interactions are
negligible. However, as the prospect of atomic BEC became more
tangible prior to 1995, Lenz {\it et al.} \cite{Lenz93,Lenz94} and
Zhang {\it et al.} \cite{ZhaWalSan94} theoretically explored the
high density regime where atom-atom interactions become relevant.
In particular, they view the cold atomic collisions as nonlinear
mixing processes analogous to those in nonlinear optics, and the
new area of nonlinear atom optics was born.  These initial
theoretical works considered the dipole-dipole interaction between
atoms, and derived a mean-field equation for the macroscopic wave
function or order parameter for the system of cold atoms analogous
to the well-known nonlinear Schr\"odinger equation from nonlinear
optics \cite{NewMol}, or the Gross--Pitaevskii equation (GPE) from
BEC theory \cite{Pita61,Gros63}. It was shown that these equations admit
soliton solutions under suitable conditions
\cite{Lenz93,ZhaWalSan94}. The experimental demonstrations of
atomic BEC in 1995 in both Rubidium \cite{Ande95} and Sodium
\cite{Davi95} ushered in the era of experimental nonlinear atom
optics, where the cold atomic collisions are now mediated via the
Van der Walls interaction, with experimental demonstrations of
atomic four-wave mixing \cite{DenHagWen99}, dark atomic solitons
\cite{Dens00,Burg99}, atomic vortices \cite{Matt99,Madi00}, and mode-locking
of
an atom laser \cite{AndKas98}.

To set the stage for our discussion, we first briefly review the
physics and derivation of the mean-field GPE for an atomic BEC,
which underlies nonlinear atom optics. We assume that an ensemble
of ultracold atoms of mass $m$ experience a single-particle
Hamiltonian of the form
\begin{equation}
H_0
=
-\frac{\hbar^2}{2m}\nabla^2
+V({\bf R}) \; ,
\end{equation}
where the external potential $V({\bf R})$ could be for example an
optical dipole potential \cite{Stam98}. In the language of
second-quantization this translates to the Hamiltonian operator
\begin{equation}
\hat{\cal{H}}_0
=
\int d^3r
\hat{\Psi}^\dagger({\bf R}) H_0 \hat{\Psi}({\bf R}) \; ,
\end{equation}
where $\hat{\Psi}^\dagger({\bf R})$ and $\hat{\Psi}({\bf R})$
denote the creation and annihilation operators (in the
Schr{\"o}dinger picture) for an atom with center-of-mass position
vector ${\bf R}$.  Assuming bosonic atoms, the field operators
obey the commutation relation $[\hat{\Psi}({\bf
R}),\hat{\Psi}^\dagger({\bf R}')] =\delta({\bf R}-{\bf R}')$. Next
we add atom-atom interactions by introducing a two-body potential
$V({\bf R},{\bf R}')$, which in second-quantization leads to the
total Hamiltonian operator ${\hat {\cal H}}={\hat {\cal
H}}_0+{\hat{\cal V}}$ with
\begin{equation}
\hat{\cal{V}}
=
\frac{1}{2}
\int d^3r d^3r'
\hat{\Psi}^\dagger({\bf R})\hat{\Psi}^\dagger({\bf R}')
V({\bf R},{\bf R}')
\hat{\Psi}({\bf R}')\hat{\Psi}({\bf R})  \; .
\end{equation}

For a system of $N$ cold atoms, the many-body state vector may be
related to the $N$-particle Schr{\"o}dinger wave function using
the general relation
\begin{equation}
\label{fockwave}
|\phi_N(t)\rangle
=
\frac{1}{\sqrt{N!}}
\int d^3r_1 \ldots d^3r_N \phi_N({\bf R}_1,\ldots,{\bf R}_N,t)
\hat{\Psi}^\dagger({\bf R}_N) \ldots \hat{\Psi}^\dagger({\bf R}_1)
|0\rangle \; .
\end{equation}
Since for a BEC at zero temperature the atoms are predominantly in
the same Bose-condensed state, we employ the Hartree approximation
in which the $N$-particle wave function $\phi_N({\bf
R}_1,\ldots,{\bf R}_N,t)$ is written in the form of a product
\begin{equation}
\phi_N({\bf R}_1,\ldots,{\bf R}_N,t)
=
\prod^N_{i=1} \vartheta({\bf R}_i,t) \; ,
\end{equation}
where $\sqrt{N}\vartheta({\bf R},t)=\langle
\phi_{N-1}|\hat{\Psi}({\bf R})|\phi_N \rangle$ is the effective
single-particle, or Hartree, wave function, and is assumed
normalized \cite{LL2}.  We note that the Hartree approximation
automatically satisfies the symmetry condition for bosons that
interchanging any two particle indices produces the same wave
function. To obtain an equation of motion for the effective
single-particle wave function, we employ Dirac's variational
principle \cite{Dirac}
\begin{equation}
\frac{\delta}{\delta \vartheta^*({\bf R},t)}
\left[
\langle\phi_N|
i\hbar\frac{\partial}{\partial t} - \left ({\hat{\cal H}}_0+
{\hat {\cal V}} \right)
|\phi_N\rangle
\right]
=
0 \;  ,
\end{equation}
which yields the GPE
\begin{equation}
\label{preGPE}
i\hbar\frac{\partial\vartheta}{\partial t}
=
H_0\vartheta
+
(N-1)\int d^3r' V({\bf R},{\bf R}')
|\vartheta({\bf R}',t)|^2\vartheta({\bf R},t) \; .
\end{equation}
Here the nonlinear term describes the mean-field
effect of the $(N-1)$ other atoms on the effective single-particle
evolution.

For the particular case of atomic BECs the cold atomic collisions
may be treated in the $s$-wave scattering limit. The effective
two-body potential can then be approximated by a contact potential
of the form \cite{LL3,Huang}
\begin{equation}
\label{delta}
V({\bf R},{\bf R}')
=
U_0\delta({\bf R}-{\bf R}') \; ,
\end{equation}
where the coefficient $U_0=4\pi\hbar^2a_{sc}/m$, with
$a_{sc}$ the $s$-wave scattering length. The GPE
equation (\ref{preGPE}) for an atomic BEC then becomes
\begin{equation}
\label{GPE}
i\hbar\frac{\partial\vartheta}{\partial t}
= -\frac{\hbar^2}{2m}\nabla^2\vartheta
+ V({\bf R})\vartheta
+
N U_0|\vartheta|^2\vartheta  \; ,
\end{equation}
which applies for large $N$.
This equation forms the basis of the following discussion of atomic
solitons.  Here we have concentrated on the case of a single atomic
species, but the formalism is straightforward to generalize to
multi-component or vector condensates \cite{Gold97,Pu98} involving, for
example,
several different Zeeman sublevels of a given atom.
%
\section{One-dimensional atomic solitons}
\label{variety}
%
\subsection{One-dimensional GPE}
In this Section we briefly discuss the one-dimensional (1D)
soliton solutions of the GPE (\ref{GPE}).  In particular, we
assume strong transverse confinement of the BEC in the $(X,Y)$ plane such
that
the transverse mode $u_g(X,Y)$ is the ground state of the
potential $V(X,Y)$,
\begin{equation}
E_gu_g = -\frac{\hbar^2}{2m} \left ( \frac{\partial^2}{\partial X^2}
+ \frac{\partial^2}{\partial Y^2} \right )u_g + V(X,Y) u_g \;  .
\end{equation}
This may be realized, for example, using a red-detuned optical
dipole trap around the focus of a Gaussian laser beam centered on
the origin \cite{Stam98}, thereby giving approximately harmonic
transverse confinement. Thus, the BEC becomes essentially
one-dimensional with cross-sectional area $A_T=\int dXdY
|u_g(X,Y)/u_g(0,0)|^2$ transverse to the unbound $Z$-axis.  We
remark that the quasi-1D nature of the system does not preclude
BEC along with the associated GPE description, as discussed by
Petrov {\it et al.} \cite{Petr00}. Setting
\begin{equation}
\sqrt{N}\vartheta({\bf R},t)=e^{-iE_gt/\hbar}
\left (\frac{u_g(X,Y)}{u_g(0,0)}\right )\varphi(Z,t) \;  ,
\end{equation}
so that $\varphi$ is normalized to the number of atoms $N$, and
projecting onto the ground transverse mode, the reduced 1D GPE
becomes (we ignore the variation of the optical potential along
the Z-axis for simplicity)
\begin{equation}
\label{GPEred}
i\hbar\frac{\partial\varphi}{\partial t}
= -\frac{\hbar^2}{2m}\frac{\partial^2\varphi}{\partial Z^2}
+ U |\varphi|^2\varphi \; ,
\end{equation}
with $U=U_0A_T^{-1}\int dXdY |u_g(X,Y)/u_g(0,0)|^4$.
\subsection{Dark solitons}
Soliton solutions are non-spreading solutions of
Eq. (\ref{GPEred}) which preserve their shape under propagation
\cite{Scott73}.
The kinetic energy term in the GPE tends to spread wavepackets,
and the nature of the solitons depends on the sign of the
nonlinearity.  For the case of repulsive interactions ($a_{sc} > 0$),
both the kinetic energy and nonlinearity terms in Eq. (\ref{GPEred})
tend to broaden localized wavepackets, so we do not expect
localized, or bright soliton, solutions for that case.
However, dark solitons describing localized density dips in an
otherwise constant background can arise and are
given by \cite{Dens00,Burg99,Rein97} (with analogous
solutions being well known in nonlinear fiber optics e.g.
\cite{Agra})
\begin{equation}
\varphi(Z,t) = n^{1/2}
\sqrt{
1-\left ( 1-\frac{v^2}{v_0^2}\right )
sech^2 \left(\frac{(Z-vt)}{l_0}\left (1-\frac{v^2}{v_0^2}\right )^{1/2}
\right )
}
e^{i(\phi(Z,t,v)-\mu t/\hbar)} \; ,
\end{equation}
where $n$ is the background density away from the dark soliton
core, $\mu=n|U|$, $l_0=\sqrt{\hbar^2/m\mu}$ is the correlation
length which determines the width of the soliton core,
$v_0=\sqrt{\mu/m}$ is the Bogoliubov speed of sound, $v$ is the
dark soliton velocity, whose magnitude is bounded by $v_0$, and
the soliton phase $\phi$ is given by
\begin{equation}
\phi(Z,t,v)=
-\arctan{\left [\left( \frac{v_0^2}{v^2}-1 \right)^{1/2}
\tanh\left(\frac{(Z-vt)}{l_0}
\left (1-\frac{v^2}{v_0^2}\right )^{1/2}
\right ) \right ]} \; .
\end{equation}
These solutions show that the dark solitons are characterized by
the presence of a phase-step $\delta$ across the localized density
dip. It can be related to the velocity $v$ and the density
$n_{bot}$ at the bottom of the atomic density dip
\cite{Dens00,Burg99}. In particular, one finds the relation
\begin{equation}
\label{phstep}
\cos\left (\frac{\delta}{2} \right ) = \frac{v}{v_0} =
\frac{n_{bot}}{n} \; ,
\end{equation}
so that for a stationary soliton $v=0, n_{bot}=0$, and there is
a $\delta=\pi$ phase-step. Only stationary solitons have a
vanishing density at their center, so they are also referred to as
`black' solitons, whereas for the $n_{bot}>0$ case the expression
`grey' soliton is used.

Dark matter-wave solitons have recently been observed
experimentally in a Na BEC by Denschlag {\it et
al.} \cite{Dens00}, and in a cigar-shaped $^{87}$Rb condensate by Burger
{\it et
al.} \cite{Dens00,Burg99}. In both experiments dark solitons of
variable velocity were launched via the phase imprinting of a BEC
by a light-shift potential. By applying a pulsed laser to only
half of the BEC and choosing the laser intensity and duration to
select a desired phase-step $\delta$, the soliton velocity could
be selected according to Eq. (\ref{phstep}). Fig. \ref{figdarkscheme} shows
a
schematic of the basic phase-imprinting idea for a cigar-shaped
BEC, and Fig. \ref{figdarkexp} shows the experimental results of Burger {\it
et
al.} The motion of the density dip is seen for a phase-step of
$\pi$ and a variety of evolution times. Ideally this value of the
phase-step should lead to a stationary dark soliton, but the
experiments are carried out in a harmonic trap \cite{Dum98} and
dissipative effects were shown to occur that accelerated the
solitons \cite{Fedi99,Busc00}.
\subsection{Bright solitons}
For the case of attractive interactions (i.e. $a_{sc} < 0$), the
kinetic energy of the BEC can be balanced by the nonlinearity
yielding non-spreading wave packets. These solutions correspond to
spatially localized bright solitons \cite{Rein97,Agra}. The
one-parameter solution to Eq. (\ref{GPEred}) then reads
\begin{equation}
\label{brightsolution}
\varphi(Z,t)
=
n^{1/2}\sqrt{2-\frac{v^2}{v_0^2}} \mbox{sech}\left[
\frac{(Z-vt)}{l_0} \left( 2-\frac{v^2}{v_0^2} \right)^{1/2}
\right] e^{i(\phi(Z,t,v)+\mu t/\hbar)} \; ,
\end{equation}
with $v$ the velocity parameter and $\mu$, $l_0$ and $v_0$ as in
the dark soliton case. The phase $\phi$ is given by
\begin{equation}
\phi(Z,t,v)
=
\frac{(Z-vt)}{l_0}\left(\frac{v}{v_0}\right) \; .
\end{equation}
Such 1D bright solitons could in principle be realized in cigar
shaped BECs with negative scattering lengths, for example, in
$^7$Li \cite{Brad95} or $^{85}$Rb by using Feshbach resonances
\cite{Corn00,Robe98,Inou98} to tune the
scattering length. To the best of our knowledge such experiments
have not been attempted.  In two or more dimensions negative
scattering lengths can lead to catastrophic collapse in
homogeneous systems for large enough particle numbers.  However,
in quasi-1D systems with strong transverse confinement the
solitons are rendered stable \cite{Kivs99}.
\subsection{Gap solitons in optical lattices}
So far in our discussion dark solitons arise for positive
scattering lengths and bright solitons for negative scattering
lengths.  However, it is possible to extend the range of options,
e.g. bright solitons with a positive scattering length, by
considering a 1D BEC in a periodic optical lattice
\cite{Berg98,Choi99}.  
This situation was first discussed by Zhang {\it et al.} \cite{Stee98}
for a scalar or single component condensate, and yields atomic gap
solitons.  Historically, gap solitons were first considered in
nonlinear optics as arising from the combination of optical
nonlinearity and a periodic spatial refractive-index distribution 
\cite{Chen87,Chri89,Ster94,Eggl96}.
As an introduction to our treatment of gap solitons in spinor
condensates, we briefly review the physics underlying gap solitons
in scalar condensates, referring the reader to the literature for
more details \cite{Stee98}.

Consider the 1D GPE (\ref{GPEred}) with an additional periodic
potential, or optical lattice, produced by periodic optical fields
and applied along the Z-axis, so that
\begin{equation}
\label{GPEopt} i\hbar\frac{\partial\varphi}{\partial t} =
-\frac{\hbar^2}{2m}\frac{\partial^2\varphi}{\partial Z^2} +
V_{opt}\cos^2(k_{opt}Z)\varphi + U |\varphi|^2\varphi \; .
\end{equation}
Here $V_{opt}$ determines the strength of the optical lattice, and
$k_{opt}$ is the wave vector characteristic of the optical
lattice.

Let us first consider the linear eigensolutions of Eq.
(\ref{GPEopt}) neglecting the nonlinear term. Introducing the 
band-index $n$ and the wave vector $k$ within the band, we can 
write the atomic wave function in terms of the periodic
Bloch functions $u_{n,k}(Z)$ according to Bloch's theorem 
\cite{Kittel}
$$ 
   \varphi(Z,t)=
   \varphi_{n,k}(Z)e^{-iE_{n,k}t/\hbar} = 
    u_{n,k}(Z)e^{i(kZ-E_{n,k}t/\hbar)} \; .
$$
Using this expression in Eq. (\ref{GPEopt}) with $U=0$, the 
eigenvalue equation associated with the cosine potential then reads
\begin{equation}
E_{n,k}\varphi_{n,k} = -\frac{\hbar^2}{2m}\frac{d^2 \varphi_{n,k}}{dZ^2}
+ V_{opt}\cos^2(k_{opt}Z)\varphi_{n,k}  \; ,
\end{equation}
the eigenvectors and eigenvalues of which 
are well known \cite{Abramowitz}. For our discussion the
important fact is the occurrence of an atomic band structure given
by the dispersion relation $E_{n,k}=\hbar\omega_{n,k}$. Consider
an atomic wavepacket for a given band-index $n$ and a narrow
spread of wave vectors $\Delta k$ around a central value $k_0$
that is slowly varying on the scale of the periodicity.
Then we can write the atomic wave function as a product of the slowly
varying envelope $\chi(Z,t)$ with the fast oscillating Bloch part
\begin{equation}
\varphi(Z,t)
=
\chi(Z,t)\varphi_{n,k_0}e^{-iE_{n,k_0}t/\hbar} \; .
\end{equation}
In this approximation the 
group velocity $v_g$ and effective mass $m_{eff}$ for the
wavepacket will be
\begin{equation}
\label{effparam}
v_g=\frac{1}{\hbar}\frac{\partial E_{n,k}}{\partial k}|_{k_0} \; ,
\qquad
\frac{1}{m_{eff}} =
\frac{1}{\hbar^2}\frac{\partial^2 E_{n,k}}{\partial k^2}|_{k_0} \; ,
\end{equation}
in terms of which the approximate dynamics of the slowly varying
envelope $\chi(Z,t)$ may be described by
\begin{equation}
\label{GPEapp}
i\hbar \left (\frac{\partial}{\partial t} +
v_g\frac{\partial}{\partial Z} \right )\chi
= -\frac{\hbar^2}{2m_{eff}}\frac{\partial^2\chi}{\partial Z^2}
+ U |\chi|^2\chi \; ,
\end{equation}
where we have re-introduced the nonlinearity.

We take for illustration the case $k_0=0$, that is zone center, so
that $v_g=0$. Then Eq. (\ref{GPEapp}) becomes identical with Eq.
(\ref{GPEred}), so that they have the same bright and dark soliton
solutions but with $m\rightarrow m_{eff}$. The key idea is that
the effective mass $m_{eff}$ can assume both positive and negative
signs by a suitable choice of the band index $n$ and $k_0$,
whereas for the spatially homogeneous case the atomic mass $m$ is
always positive.  So now, for example, bright solitons can arise
for a positive scattering length if the effective mass is
negative, the general condition for bright solitons being
$m_{eff}U<0$, and conversely for dark solitons $m_{eff}U>0$.  
We remark that the applied optical lattice can
modify the effective mass along the $Z$-axis, but the effective
mass along the transverse dimensions is still $m$. This leaves
open the question of higher-dimensional gap solitons, a point that
we do not further address here \cite{Akoz98}.

The solitons formed in optical lattices are referred to as gap
solitons \cite{Chri89,Ster94,Eggl96}.  This name derives from the
fact that the
soliton energies actually lie in the energy gaps of the atomic
band structure \cite{Chen87}.  The significance of gap solitons in
the
context of atomic BEC is that they provide a means to realize
bright solitons, that is, spatially localized atomic packets, 
{\em even for positive scattering lengths.} This may have important
applications, e.g. for the coherent transport of atoms. In the
following section we further develop the theory of atomic gap
solitons in spinor condensates to quantify some of their most
important properties.
%
\section{Atomic gap solitons}
\label{lattices}
%
We now consider a system of two condensates in different Zeeman
sublevels coupled by a spatially periodic two-photon interaction.
This spatially modulated coupling again leads to an atomic band
structure for the spinor condensate, and hence by our general
arguments of the previous section, to gap solitons.  The advantage
of multicomponent condensates is that magnetic fields can now be
used to phase-imprint the two Zeeman components differently, and
hence to manipulate and control the gap solitons to a high degree.
We demonstrate this flexibility in Section V by simulating an
atomic interferometer based on gap solitons.
\subsection{The physical model for a spinor condensate}
The two-component Bose--Einstein condensate interacts with two
counterpropagating, focused Gaussian laser beams of equal
frequencies $\omega_l$ but opposite circular polarizations. The
optical dipole potential associated with the applied laser beams
is assumed to provide tight transverse confinement for the BEC in
the $(X,Y)$ plane, thereby forming a cigar-shaped condensate of
transverse cross-sectional area $A_T$. As in Section
\ref{variety}, we confine our discussion to the one-dimensional
dynamics of the BEC along the $Z$-axis for simplicity.

In addition to supplying a transverse optical potential, the laser
beams can drive two-photon transitions between different Zeeman
sublevels of the atomic ground state. For illustrative purposes we
consider the case of Sodium and the two-photon coupling of the
Zeeman sublevels $|-1\rangle=| F_g = 1, M_g=-1\rangle$ and
$|1\rangle=| F_g = 1, M_g=1\rangle$. For example, starting in the
$|-1\rangle$ state this process involves the absorption of a
$\sigma_+$ photon from the right propagating laser beam followed
by emission of a $\sigma_-$ photon into the left propagating laser
beam. We must of course assume that the excited states involved in the
atom-field interaction are far-detuned from the applied laser
frequency, a necessary requirement to avoid the detrimental
effects of spontaneous emission.

By restricting our attention to the coupled states $|\pm 1\rangle$
the effective single-particle Hamiltonian for our model system
can be written as \cite{Zoba99,Cohe92}
\begin{equation}
\label{eq:heff} H_{\mathit{eff}} = \frac{P_Z^2}{2m}+g\hbar\delta'
                 \left[
                    \left|1\right\rangle\left\langle -1\right| e^{2iK_l Z} +
                    \left|-1\right\rangle\left\langle 1\right| e^{-2iK_l Z}
                 \right ] \; ,
\end{equation}
where we have omitted constant light-shift terms.  Here $P_Z$ is
the atomic center-of-mass momentum operator along $Z$,
$K_l=\omega_l/c$ is the magnitude of the field wave vector along
$Z$, $g$ is a coupling constant between the ground and excited
states characteristic of the atom and transition involved,
$\delta' = {\mathcal{D}}^2{\mathcal{E}}^2/\hbar^2 \delta$, with
the detuning $\delta = \omega_l - \omega_a$, $\mathcal{E}$ is the
laser field amplitude, and $\mathcal{D}$ is the reduced electric
dipole moment for the $3S_{1/2}$-$3P_{3/2}$ transition.

The second term of the effective Hamiltonian (\ref{eq:heff})
describes the effective coupling of the two Zeeman sublevels via
the applied laser fields. The exponential terms $\exp(\pm 2iK_lZ)$
arise from the fact that the two-photon transitions involve the
absorption of a photon from one light field and reemission into
the other. Introducing a spinor macroscopic condensate wave
function $\varphi(Z,t) =[\varphi_{1}(Z,t),\varphi_{-1}(Z,t)]^T$
normalized to the number of atoms $N$, and including the
many-body effects via a mean-field nonlinearity, we obtain the
coupled GPE equations
\begin{equation}
\label{gpcop} i\hbar\frac{\partial\varphi}{\partial t}
=H_{\mathit{eff}}\varphi + U|\varphi|^2\varphi \; ,
\end{equation}
where $U=4\pi\hbar^2a_{sc}/m$ as in Section \ref{NLAO} and \ref{variety},
$|\varphi|^2=|\varphi_1|^2+|\varphi_{-1}|^2$, and we have assumed
that the magnitude of the self- and cross-nonlinearities are equal
for simplicity.

It is convenient to re-express Eq.\ (\ref{gpcop}) in dimensionless
form by introducing the scaled variables $t=\tau t_c$, $Z =z
l_{c}$ and $\varphi_j=\psi_j\sqrt{\rho_{c}}$ where
\begin{equation}
  \label{eq:scalefactors}
  t_c = \frac{1}{g \delta'} \; , \quad
  l_c = \frac{t_c \hbar K_l}{m} \; , \quad
  \rho_c = \left|\frac{g\hbar\delta'}{U}\right| \; .
\end{equation}
Equations (\ref{gpcop}) then become
\begin{eqnarray}
  \label{eq:scaled}
  i \frac{\partial}{\partial \tau}
  \left(\begin{array}{c} \psi_1 \\ \psi_{-1} \end{array}\right)
  & = &
  \left(\begin{array}{cc}
        \displaystyle{-M\nabla^2} &
        \displaystyle{e^{2ik_lz}} \\
        \displaystyle{e^{-2ik_lz}} &
        \displaystyle{-M\nabla^2}
  \end{array}\right)
  \left(\begin{array}{c} \psi_1 \\ \psi_{-1} \end{array}\right) \nonumber\\
  & + & \mbox{sgn}\left(g\delta'/U\right)\left|\psi\right|^2
  \left(\begin{array}{c} \psi_1 \\ \psi_{-1} \end{array}\right) \; ,
\end{eqnarray}
where $M=g \delta' m/2 \hbar K_l^2$ is a mass-related parameter
such that $k_l = K_l l_c = 1/2M$. For a discussion of characteristic
values for the case of a Sodium condensate we refer the reader to Ref.
\cite{Pott00}.

\subsection{Soliton solutions}
\begin{figure}
\includegraphics*[width=0.8\columnwidth]{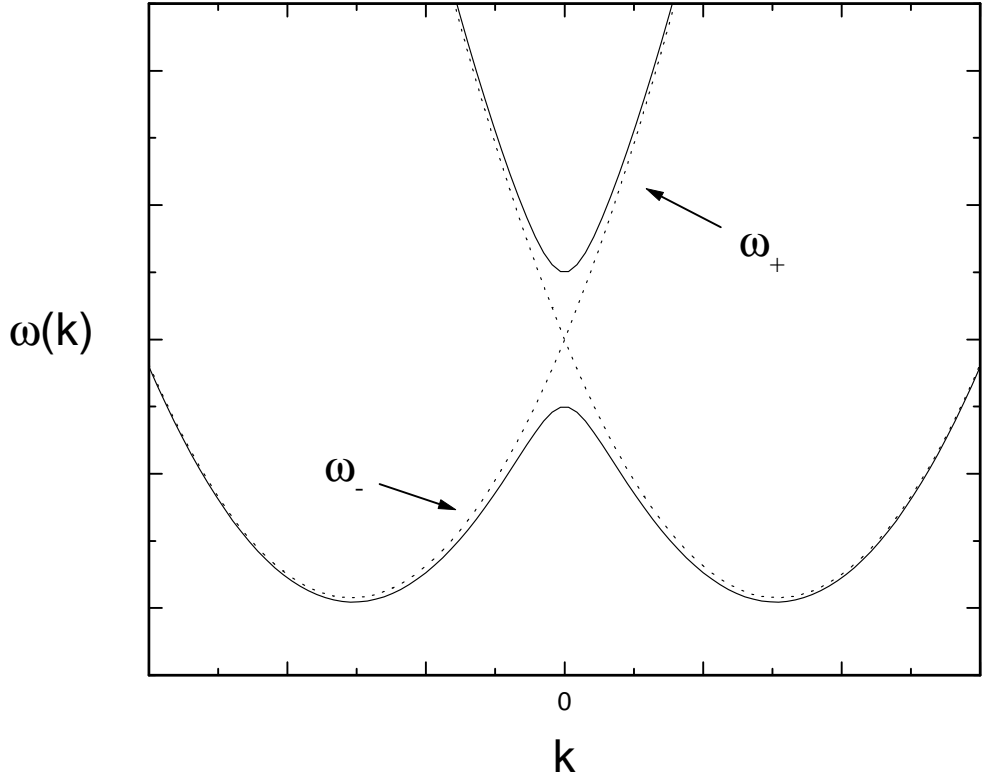}
\caption{Linear dispersion curve (arbitrary units): Shown are the two
         branches of
         the dispersion relation exhibiting an avoided crossing at
         $k=0$ ({\em solid line}). The negative curvature in the lower branch around
         $k=0$ defines a negative effective mass that enables
         gap soliton solutions. The free dispersion when no linear 
         coupling is present corresponds to two displaced parabolas
         ({\em dashed line})}
\label{figdispersion}
\end{figure}

The spatially modulated coupling between the optical fields and
the condensate induces a single-particle band structure with
regions of negative effective mass. As mentioned before, this
leads to the possibility of bright atomic solitons even for
repulsive interactions \cite{Zoba99}. Approximate analytic
expressions for these solitons can be obtained by expressing the
spinor condensate components as
\begin{equation}
  \label{eq:factoroutsvea}
  \psi_{\pm1}(z,\tau)
  =
  e^{\pm ik_lz}e^{-i\tau/4M}\phi_{\pm1}(z,\tau) \; ,
\end{equation}
where the field envelopes $\phi_{\pm1}$ are assumed to be slowly
varying in space compared to $1/k_l$.  Neglecting the second-order
spatial derivatives yields the coupled partial differential
equations
\begin{eqnarray}\label{gapp}
i\left (\frac{\partial}{\partial\tau}\pm \frac{\partial}{\partial
z} \right ) &&\left(\begin{array}{c} \phi_1
\\ \phi_{-1}\end{array} \right)=\left(\begin{array}{cc} 0 & 1\\ 1
& 0 \end{array}\right) \left(\begin{array}{c} \phi_1 \\ \phi_{-1}
\end{array} \right) \nonumber \\ &&\pm (|\phi_1|^2+|\phi_{-1}|^2)
\left(\begin{array}{c} \phi_1 \\ \phi_{-1}
\end{array} \right) \;  ,
\end{eqnarray}
where the choice $\pm 1=\mbox{sgn}(g\delta^\prime/U)$. For a
red-detuned laser and our choice of $g$ this becomes $\pm
1=\mbox{sgn}(U)$. Neglecting the nonlinearity one finds a linear
dispersion curve as sketched in Fig. \ref{figdispersion}. In this
particular case it is the linear coupling induced by the laser
field that creates the avoided crossing at $k=0$ and the negative
effective mass. Aceves and Wabnitz \cite{Acev89} have shown that
the full equations (\ref{gapp}) have the explicit two-parameter
gap soliton solutions (see also Ref. \cite{Ster94})
\begin{eqnarray}
\phi _{1}&=&\pm \frac{\sin(\eta)}{\beta\gamma\sqrt{2}}
            \left( -\frac{e^{2\theta }+e^{\mp i\eta }}
             {e^{2\theta }+e^{\pm i\eta }}\right) ^{v}sech
            \left( \theta \mp \frac{i\eta }{2}\right)
            e^{\pm i\sigma } \; ,
\nonumber \\ \phi_{-1}&=&-\frac{\beta\sin(\eta)}{\gamma\sqrt{2}}
             \left( -\frac{e^{2\theta }+e^{\mp i\eta }}
              {e^{2\theta }+e^{\pm i\eta }}\right) ^{v}sech
             \left( \theta \pm \frac{i\eta }{2}\right)
             e^{\pm i\sigma } \; .
\label{GapEnv}
\end{eqnarray}
Here $-1<v<1$ is a parameter which controls the soliton velocity,
$0<\eta<\pi$ is a shape parameter, and

\begin{equation}
\label{eq:param1} 
\beta =\left( \frac{1-v}{1+v}\right)^{\frac{1}{4}} \; , 
\quad 
\gamma =\frac{1}{\sqrt{1-v^{2}}} \; ,
\end{equation}
\begin{equation}
\label{eq:param2} 
\theta =-\gamma\sin(\eta)(z-v\tau ) \; , 
\quad
\sigma =-\gamma\cos(\eta)(vz-\tau ) \;.
\end{equation}
Since we are interested in creating bright solitons in the
presence of repulsive interactions we restrict ourselves to
$\mbox{sgn}(U)=+1$, corresponding to the choice of the upper sign
in the analytic solutions. The characteristic length scale
associated with the solitons is $l_c$, so that the approximate
solitons (\ref{GapEnv}) are valid for $K_ll_c = 1/2M >> 1$.

\subsection{Soliton properties}
From the dependence of the hyperbolic-secant on
$\theta=-\gamma\sin(\eta)(z-v\tau)$ in Eqs. (\ref{GapEnv}), we
identify the gap soliton parameter $v=V_g/V_R$ as the group
velocity $V_g$ of the soliton in units of the recoil velocity
$V_R=l_c/t_c=\hbar K_l/m $. Since $-1<v<1$, the magnitude of the
group velocity is bounded by the recoil velocity. From Eqs.
(\ref{GapEnv}), one can extract further important soliton
properties, such as the number $N_s$ of atoms in the soliton and
the soliton width $W_s=w_sl_c$. Specifically, the number of atoms
in a particular gap soliton is given by
\begin{equation}
N_s=A_T\int dZ [|\varphi_1(Z,t)|^2 + |\varphi_{-1}(Z,t)|^2] \; , \label{Ns}
\end{equation}
where $A_T$ is the effective transverse area.

We can gain further insight into the analytical structure of the
gap solitons by taking the extreme limit $\eta<<1$ of Eqs.
(\ref{GapEnv}). Using the definitions (\ref{eq:factoroutsvea}) and
returning to dimensional units, we have then
\begin{eqnarray}
\varphi_{1}(Z,0)&=&
            \frac{\eta}{\beta\gamma}\sqrt{\frac{\rho_c}{2}}
            sech(Z/W_0)(-1)^v e^{i(K+K_l)Z}  \; ,
\nonumber \\ \varphi_{-1}(Z,0)&=&
\frac{\beta\eta}{\gamma}\sqrt{\frac{\rho_c}{2}}
             sech(Z/W_0)(-1)^v e^{i[(K-K_l)Z + \pi]} \; ,
\label{GapTot}
\end{eqnarray}
where
\begin{equation}
W_s = 3.44W_0=3.44\left (\frac{l_c\sqrt{1-v^2}}{\eta}\right ) \; ,
\quad K=-\frac{\gamma v}{l_c} \; . 
\label{SolPar}
\end{equation}
Here $W_s=3.44W_0$ is the soliton width, the factor $3.44$ being
the numerical conversion from the width of the hyperbolic secant
to the $1/e^2$ width of the distribution, and $K=k/l_c$ a
velocity-dependent wave  vector shift. Obviously the width
decreases with increasing $\eta$ and $v$. The soliton atom number
obtained by combining Eqs. (\ref{Ns}) and (\ref{GapTot}),
\begin{equation}
N_s = 2(A_Tl_c\rho_c)\eta\cdot(1-v^2) \; ,
\end{equation}
predicts that increasing the shape parameter $\eta$ increases the atom
number,
and faster solitons have lower atom numbers.

An essential point to keep in mind is that the gap solitons are
coherent superpositions of the two Zeeman sublevels. The
approximate solutions (\ref{GapTot}) contain important information
on the phase and amplitude relations that need to be created
between them to successfully excite and manipulate gap solitons.
In particular, they show that there is always a spatially
homogeneous $\pi$ phase difference between the two states.  In
addition, the two components have the spatial wave vectors
\begin{equation}
K_{\pm 1} = K \pm K_l  \; , 
\label{Kpm}
\end{equation}
with $K=-\gamma v/l_c$.

Finally, it follows from dividing
the amplitudes of the two components that
\begin{equation}
\left |\frac{\varphi_1(Z,t)}{\varphi_{-1}(Z,t)} \right |^2 =
\frac{1}{\beta^4} = \left (\frac{1+v}{1-v} \right ) \; , 
\label{Rpm}
\end{equation}
which shows that their relative occupation depends on the soliton
velocity parameter $v$. For $v=0$ the sublevels are equally
populated, but as $v\rightarrow 1$ the $|1\rangle$ sublevel has a larger
population, and vice versa for $v\rightarrow -1$.

The characteristic time scale for the evolution of the gap
solitons can be determined from the plane-wave exponential factors
in Eqs. (\ref{GapEnv}). Converting back to dimensional form the
soliton period $t_s$ is defined as the time to accumulate a $2\pi$
phase, or in the limit $\eta\rightarrow 0$
\begin{equation}
t_s = 2\pi t_c\sqrt{1-v^2} \; .
\label{solper}
\end{equation}
Physically, $t_s$ corresponds to the internal time scale for the
gap soliton. In order to observe a soliton-like behavior, it is
therefore necessary to investigate the atomic propagation over
several periods.

We conclude this section by noting that Aceves and Wabnitz
\cite{Acev89} have shown that the gap soliton solutions
(\ref{GapEnv}) are stable solutions of Eqs. (\ref{gapp}) in that
they remain intact during propagation, even when perturbed away
from the exact solutions. However, one should remember that Eq.
(\ref{gapp}) is only an approximation to the exact system of
equations (\ref{eq:scaled}), so that in general the gap solitons
are solitary wave solutions only. As such, they are not guaranteed
to be absolutely stable.

\section{Magneto-optical control}
\label{control}

Summarizing the main results of the preceding section, we have
seen that gap solitons require the right population and density
distribution in each Zeeman sublevel, a certain phase difference
between these sublevels, and appropriate plane-wave factors
$e^{iK_{\pm1} Z}$. Based on these properties, we now show that a
magneto-optical scheme involving a combination of pulsed coherent
optical coupling and of phase-imprinting using spatially
inhomogeneous magnetic fields turn out to be an adequate and
convenient tool to manipulate these solitons.

\subsection{Manipulation Tools}
\label{tools}

The shapes of the Hartree wave functions corresponding to the two
Zeeman sublevels are hyperbolic-secant, which we approximate by a
Gaussian in the following. They could for example be initialized
in an optical dipole trap \cite{Stam98}. Manipulating the gap
solitons reduces therefore to the problem of controlling the
populations and phases throughout the spinor condensate. The
coherent optical coupling can be achieved e.g. by a laser pulse of
frequency $\omega_l$ propagating perpendicularly to the $Z$-axis
and with linear polarization perpendicular to that axis. For
sufficiently short pulses, one can neglect changes in the
center-of-mass motion of the atoms during its duration, leading to
a very simple description. We assume for simplicity a plane-wave
rectangular pulse of duration $t_p$ and of spatial extent large
compared to the soliton. The Hamiltonian describing the coupling
between this pulse and the condensate is then the same as in Eq.
(\ref{eq:heff}), but without the linear momentum exchange terms
$\exp(\pm 2iK_l Z)$ and the kinetic energy term, and with
$\delta^\prime \rightarrow\delta_p^\prime$. The state of the
system after the pulse is then easily found to be
\begin{eqnarray}
\label{eq:pulsesolution} \left(\begin{array}{c} \varphi_1(t_p) \\
\varphi_{-1}(t_p)\end{array}\right) & = & \left(\begin{array}{cc}
          \displaystyle{\cos\chi} &
          \displaystyle{i\sin\chi} \\
          \displaystyle{-i\sin\chi} &
          \displaystyle{\cos\chi}
\end{array}\right)
\left(\begin{array}{c} \varphi_1(0) \\ \varphi_{-1}(0)\end{array}\right)
\nonumber \\ & \equiv & M_L(\chi) \left(\begin{array}{c} \varphi_1(0)
\\ \varphi_{-1}(0)\end{array}\right) \; .
\end{eqnarray}
where $\chi=g\delta_p^\prime t_p$ is the excitation pulse area and
the operator $M_L$ can be used to control the population transfer
by an appropriate choice of $\chi$.

The required phase relationship between the two states can be
achieved via Zeeman splitting. Considering for concreteness a
spatially inhomogeneous rectangular magnetic field pulse of
duration $t_B$ we have, neglecting again all other effects,
\begin{equation}
\label{eq:bfieldequation}
i\hbar\frac{\partial}{\partial t}
\varphi_{\pm1}(Z,t)
=
\pm \mu_B g_F \left(B_0+B'Z\right)\varphi_{\pm1}(Z,t) ; ,
\end{equation}
where $g_F$ is the Land\'e g-factor of the hyperfine ground state,
$\mu_B$ is the Bohr magneton, $B_0$ the spatially homogeneous
component of the magnetic field, and $B'$ its gradient, the
direction of the magnetic field being along the $Z$-axis. The
application of this field results in the state
\begin{eqnarray}
\label{eq:bfieldsolution} \left(\begin{array}{c} \varphi_1(t_B)
\\ \varphi_{-1}(t_B)\end{array}\right) & = &
\left(\begin{array}{cc}
          \displaystyle{e^{i(\vartheta + K_B Z)}} & \displaystyle{0} \\
          \displaystyle{0} &
          \displaystyle{e^{-i(\vartheta + K_B Z)}}
\end{array}\right)
\left(\begin{array}{c} \varphi_1(0) \\ \varphi_{-1}(0)\end{array}\right)
\nonumber \\ & \equiv & M_B(\vartheta,K_B) \left(\begin{array}{c}
\varphi_1(0) \\ \varphi_{-1}(0)\end{array}\right) \; ,
\end{eqnarray}
where $\vartheta = -(\mu_B g_F/\hbar)B_0t_B$ and $K_B = -(\mu_B
g_F/\hbar)B't_B$ are the imprinted phase shift and phase gradient
(or wave vector), respectively. That is, the application of the
magnetic pulse results in a phase difference of $2\vartheta$
between the two Zeeman sublevels, and in addition it imparts them
wave vectors $\pm K_B$.

\subsection{Excitation and Application}

\begin{figure}
\includegraphics*[width=0.8\columnwidth]{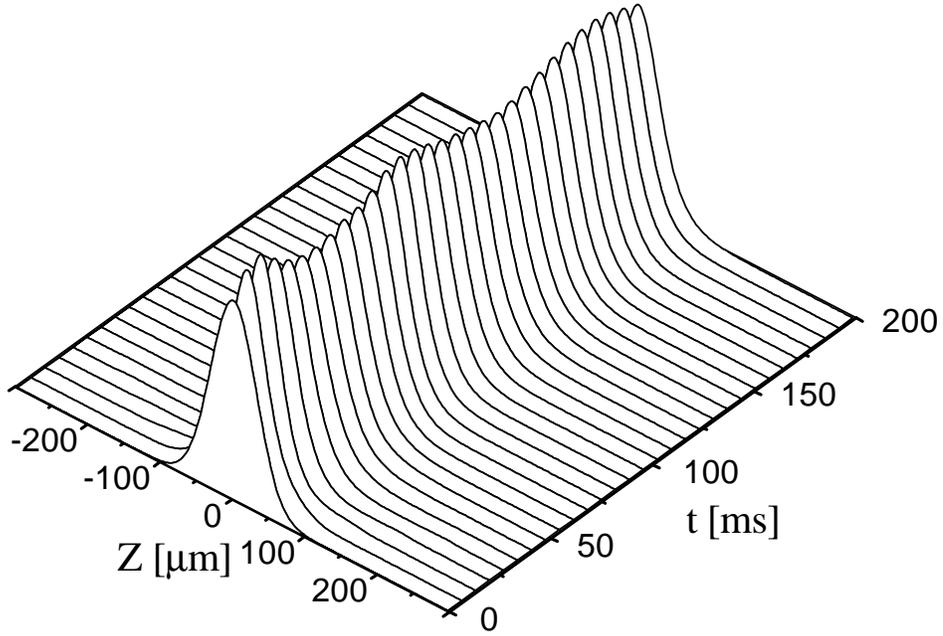}
\caption{The propagation of the total density of a stationary soliton
         $(v=0)$ over around 50 soliton periods.
         Peak density oscillations are due to imperfect initial conditions}
\label{figexcitation}
\end{figure}

To illustrate how stationary and moving solitons can be excited
using this scheme, we start from a scalar condensate in the
$|-1\rangle$ state, $\varphi(Z,0)=[0,\varphi_0(z)]^T$, with
spatial mode
\begin{equation}
\label{init} \varphi_0(Z)=\frac{N_s}{\sqrt{A_T}}\left (\frac{2}{\pi
W_s^2} \right )^{1/4} e^{-Z^2/W_s^2} \; ,
\end{equation}
with $W_s$ and $N_s$ the width and atom number of the gap soliton
desired. This Gaussian is chosen to approximate the
hyperbolic-secant structure of the analytic gap soliton solution
in Eq. (\ref{GapTot}).  For a stationary solution we need to
prepare the Zeeman sublevels with equal populations and with a
$\pi$ phase difference. We further need to impose wave vectors
which are equal in magnitude but opposite in sign, $K_{\pm 1}=\pm
K_l$. This can be achieved by applying a laser pulse of area
$\chi=\pi/4$, followed by a magnetic pulse with $\vartheta=\pi/4$
and $K_B=K_l$. The state then transforms as $(t=t_p+t_B)$
\begin{eqnarray}
\label{eq:excitestationary} \varphi(Z,t) & = &
M_B(\pi/4,K_l)M_L(\pi/4)\varphi(Z,0) \nonumber \\ & = &
\frac{e^{\frac{3i\pi}{4}}} {\sqrt{2}}\left(\begin{array}{c}
e^{iK_l Z}\\ e^{-i\pi}e^{-i K_l Z}\end{array}\right)\varphi_0(Z) \; .
\end{eqnarray}
Figure \ref{figexcitation} shows the resulting stable evolution of
the total density $\left|\varphi(Z,t)\right|^2$. As a result of
the  Gaussian approximation to the exact solution there are some
slow oscillations imposed on the motion, but the solution remains
centered at $Z=0$ and stationary over a time $t=200$ ms, much
longer than the soliton period for the chosen values. We note that
although Section \ref{tools} treated the pulsed excitations in an
impulsive manner to intuitively understand their action, our
simulations do not make this approximation. The numerics confirm
the accuracy of the impulsive approximation for the parameters at
hand; a consequence of the fact that we consider pulse durations
significantly shorter than the soliton period (\ref{solper}).

The excitation of a moving soliton is slightly more complicated
since the velocity dependent wave vector $K=-\gamma v/l_c$ in Eq.
(\ref{GapTot}) is no longer zero and the two components have
different populations, see Eq. (\ref{Rpm}). However, our
simulations show that the proposed scheme performs well on this
task, as well as on splitting solitons and reversing their
directions \cite{Pott00}.

Solitons present some advantages for atom interferometry in that
they are many-atom wavepackets which are immune to the effects of
spreading, hence allowing longer path lengths, and also increased
signal-to-noise for large atom numbers. Typically many-body
effects limit the utility of high-density wavepackets due to
spatially varying mean-field phase shifts, but solitons have the
cardinal virtue that they have fixed spatial phase variations.
Thus they may provide a key to making maximal use of high density
sources for atom interferometry. Indeed, due to these very
properties they have long been advocated for all-optical switching
applications.

\section{Summary}
\label{summary}

In this paper, we have reviewed important aspects of the
matter-wave solitons that can be launched in weakly interacting
Bose--Einstein condensates. So far, only dark solitons have been
demonstrated experimentally, but the rapid progress in the optical
manipulation and control of condensates is expected to lead to the
demonstration of bright gap solitons.

\begin{figure}
\includegraphics*[width=0.8\columnwidth]{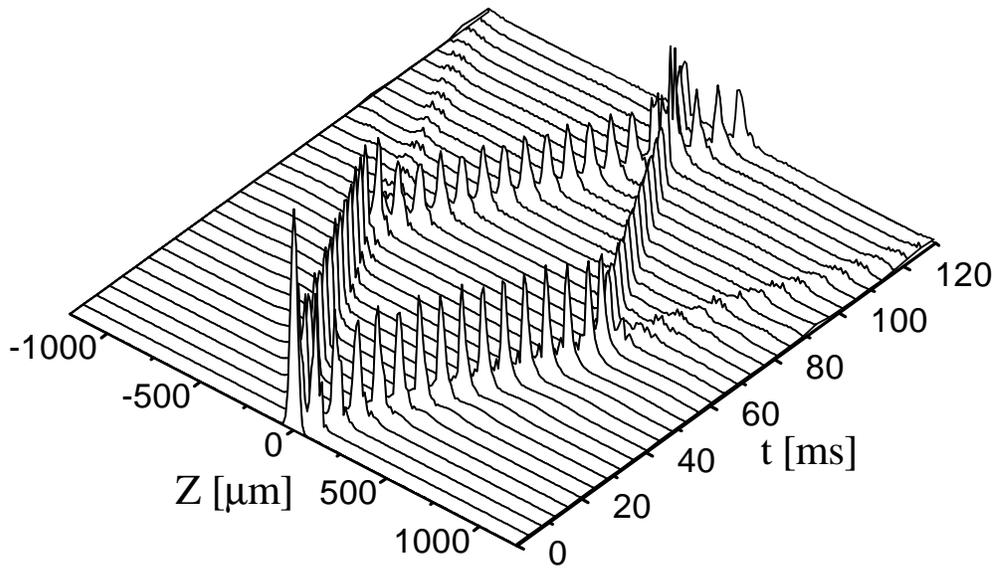}
\caption{Demonstration of an atomic Mach--Zehnder interferometer: The
         initial
         condensate is split into two counterpropagating
         solitons, then their direction is reversed and they collide
         (shown is the total density)}
\label{figinterferometer}
\end{figure}

\begin{figure}
\includegraphics*[width=0.8\columnwidth]{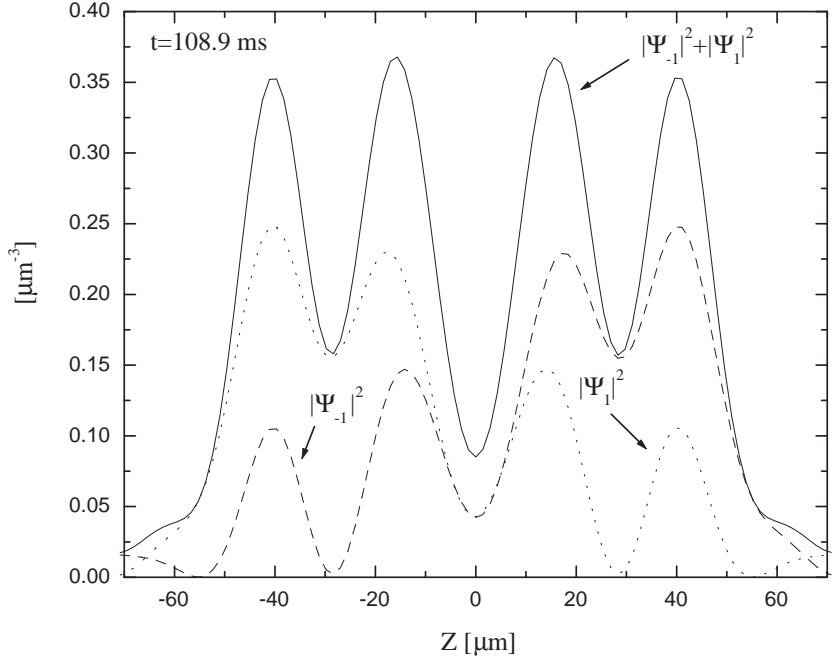}
\caption{Interference pattern when two solitons collide in the
         Mach--Zehnder configuration: The
         total density ({\em solid line}) is symmetric, whereas the
         interference pattern of each of the two orthogonal Zeeman
         states ({\em dotted} and {\em dashed line}) is
         asymmetric due to the shape of the colliding wavepackets. The
         contrast in the total density pattern is around 30\%}
\label{figinterference}
\end{figure}

As an illustration of the potential use of these objects, we
consider a soliton-based nonlinear atomic Mach--Zehnder
interferometer. Our specific demonstration involves an initial
scalar condensate that is split into two oppositely moving
solitons along $Z$, see Fig. \ref{figinterferometer} for $t<60$
ms.  At $t=60$ ms laser and magnetic pulses are applied which act
to reverse the direction of the two solitons. This process causes
some loss of atoms in both solitons, as before.  The two reversed
soliton components come together again at $t=120$ ms.  Since the
colliding solitons are predominantly in opposite orthogonal Zeeman
sublevels, the interference pattern appearing during the collision
is due to the contamination of each soliton by the other state.
Fig. \ref{figinterference} shows the interference at the soliton
collision in the total density (solid line) and also the
individual Zeeman sublevels, with a fringe contrast of around
30\%. These results demonstrate the potential use of gap solitons
for realizing nonlinear atom interferometers with high brightness
sources.

\begin{figure}
\includegraphics*[width=0.8\columnwidth]{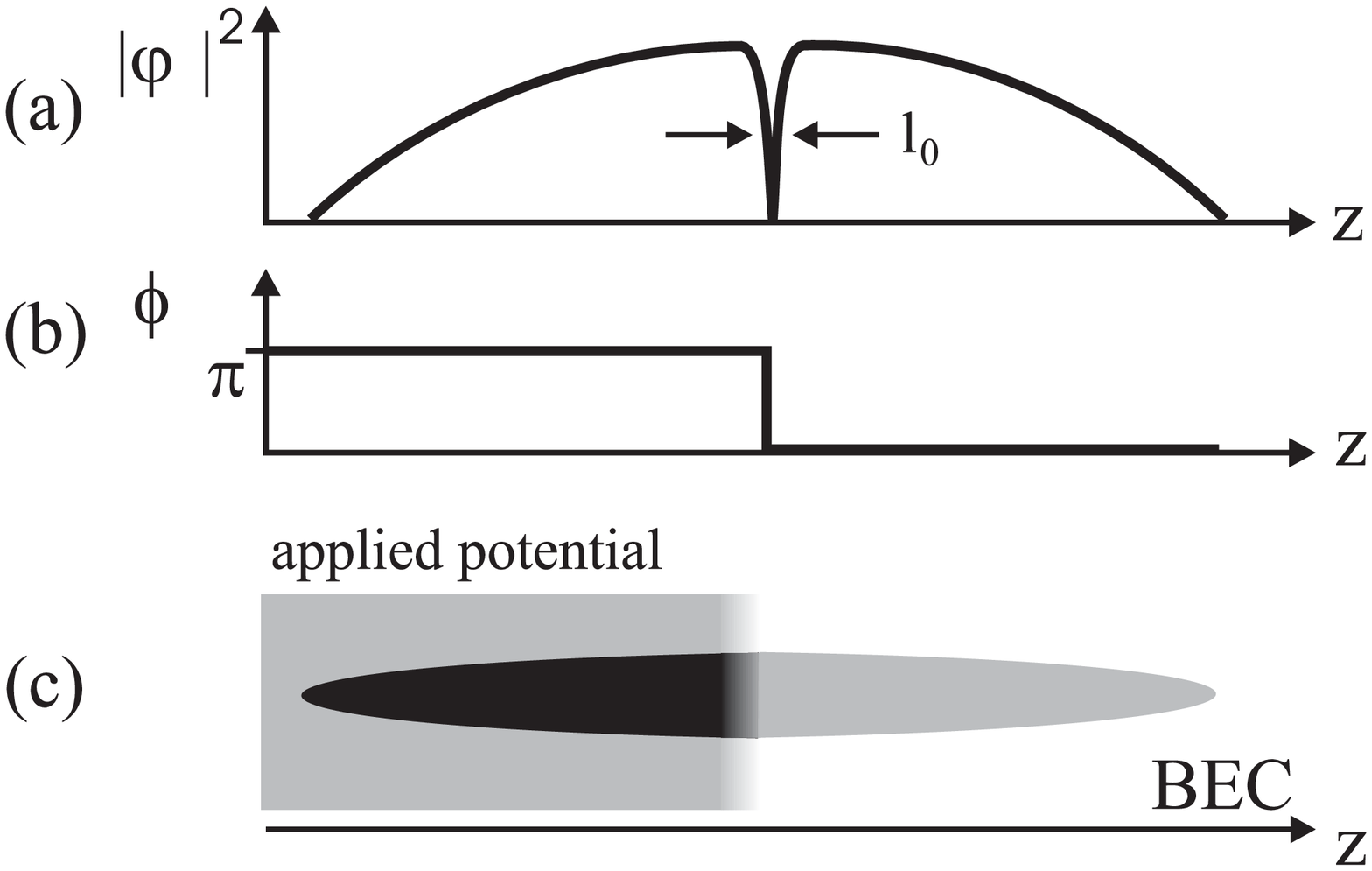}
\caption{Density distribution ({\bf a}) and spatial phase ({\bf b}) of a
  stationary dark soliton with $\delta=\pi$. The dip in the density has
  a width $\sim l_0$. The scheme for the generation of dark solitons by
  phase imprinting is shown in ({\bf c})}
\label{figdarkscheme}
\end{figure}

\begin{figure}
\includegraphics*[width=0.8\columnwidth]{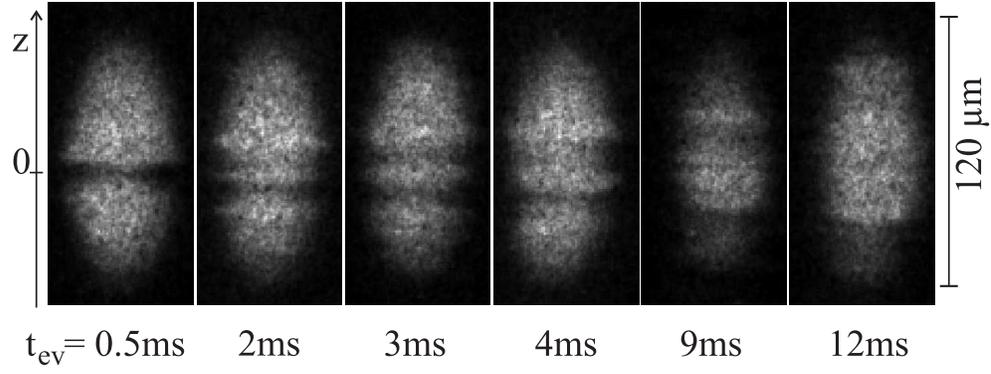}
\caption{Absorption images of BECs with dark soliton structures
  propagating along the long condensate axis for different evolution
  times $t_{ev}$}
\label{figdarkexp}
\end{figure}

\acknowledgments 
The authors would like to thank the Hannover group 
for providing Figs. \ref{figdarkscheme} and \ref{figdarkexp} as
well as K.~M. Hilligsoe for poiting out a mistake in (\ref{effparam}).
This work is supported in part by Office of Naval
Research Contract Nos.~14-91-J1205 and N00014-99-1-0806, 
the National Science Foundation
Grant PHY98-01099, the Army Research Office and the Joint Services
Optics Program.

\newpage

\newpage

\newpage

\newpage

\newpage

\newpage

\end{document}